\documentclass[5p,twocolumn,authoryear,final]{elsarticleaccepted}
\usepackage{amsmath}
\usepackage{amssymb}
\usepackage{txfonts}
\DeclareSymbolFont{cmletters}{OML}{cmm}{m}{it}
\DeclareMathSymbol{v}{\mathord}{cmletters}{"76}

\usepackage{hyperref}

\newcommand{\useiop}{-2}

\newcommand\figwidthfactor{1}
\newcommand{\shortauthors}[1]{}
\newcommand{\shorttitle}[1]{}
\newcommand{\acknowledgements}{\section*{Acknowledgements}}
\newcommand{\altaffiltext}[2]{\address{#2}}
\newcommand\apj{\rmfamily{ApJ}}%
\newcommand\apjl{\rmfamily{ApJ}}%
\newcommand\apjs{\rmfamily{ApJS}}%
\newcommand\mnras{\rmfamily{MNRAS}}%
\usepackage{ifthen}
\usepackage[usenames,dvipsnames]{color}

\defcitealias{tch08}{TMN08}
\defcitealias{tch09}{TMN09}
\defcitealias{kom09}{K09}

\newcommand{\abs}[1]{\ensuremath{\left|#1\right|}}

\newcommand{\cut}[1]{\hbox{}}

\shortauthors{A. Tchekhovskoy, R. Narayan, \& J.~C. McKinney}
\shorttitle{Magnetohydrodynamic Simulations of Gamma-Ray Burst Jets:
  Beyond the Progenitor Star}

\begin{document}
\label{firstpage}

\ifthenelse{\equal{\useiop}{-2}}{
\journal{New Astronomy}
\begin{frontmatter}

\title{Magnetohydrodynamic simulations of gamma-ray burst jets: beyond
  the progenitor star}

\author[cfa]{Alexander Tchekhovskoy}
\ead{atchekho@cfa.harvard.edu}

\author[cfa]{Ramesh Narayan}
\ead{rnarayan@cfa.harvard.edu}

\author[stanford]{Jonathan C. McKinney}
\ead{jmckinne@stanford.edu}

\address[cfa]{Institute for Theory and Computation,
  Harvard-Smithsonian Center for Astrophysics, 60 Garden Street,
  Cambridge, MA 02138, USA}
\address[stanford]{Kavli Institute for
  Particle Astrophysics and Cosmology, Stanford University, P.O. Box
  20450, Stanford, CA 94309, USA; Chandra Fellow}
}
{
\title{Magnetohydrodynamic Simulations of Gamma-Ray Burst Jets: Beyond the Progenitor Star}
\author{Alexander Tchekhovskoy,$^1$ Ramesh
  Narayan$^1$, Jonathan C. McKinney$^2$}
\altaffiltext{1}{Institute for Theory and Computation,
  Harvard-Smithsonian Center for Astrophysics, 60 Garden Street,
  Cambridge, MA 02138, USA; atchekho@cfa.harvard.edu,
  rnarayan@cfa.harvard.edu}\altaffiltext{2}{Kavli Institute for
  Particle Astrophysics and Cosmology, Stanford University, P.O. Box
  20450, Stanford, CA 94309, USA; Chandra Fellow;
  jmckinne@stanford.edu}
}

\begin{abstract}
  Achromatic breaks in afterglow light curves of gamma-ray bursts
  (GRBs) arise naturally if
  the product of the jet's Lorentz factor
  $\gamma$ and opening angle $\Theta_j$ satisfies
  $\gamma\Theta_j\gg1$ at the onset of the afterglow
  phase, i.e., soon after the conclusion of the prompt emission.
  Magnetohydrodynamic (MHD) simulations of
  collimated GRB jets generally give $\gamma\Theta_j\lesssim1$,
  suggesting that MHD models may be inconsistent with jet breaks.
  We work within the collapsar paradigm and use axisymmetric
  relativistic MHD simulations to explore the effect of a finite
  stellar envelope on the structure of the jet.  Our idealized models
  treat the jet--envelope interface as a collimating rigid
  wall, which opens up outside the star to mimic loss of collimation.
  We find that the onset of deconfinement causes a
  burst of acceleration accompanied by a slight increase in the opening angle.
  In our fiducial model with a stellar radius equal to $10^{4.5}$
  times that of the central compact object, the jet achieves an
  asymptotic Lorentz factor $\gamma\sim500$ far outside the star and
  an asymptotic opening angle
  $\Theta_j\simeq0.04\thickspace{}{\rm{}rad}\medspace\simeq2^\circ$,
  giving $\gamma\Theta_j\sim20$.  These values are consistent with
  observations of typical long-duration GRBs, and explain the
  occurrence of jet breaks.  We provide approximate analytic solutions
  that describe the numerical results well.

\end{abstract}

\ifthenelse{\equal{\useiop}{-2}}{
\end{frontmatter}
}{
}

%%%%%%%%%%%%%%%%%%%%%%%%%%%%%%%%%%%%

\section{Introduction}
\label{sec_intro}

Relativistic jets are ubiquitous features of many accreting black
holes and neutron stars.  They are found in systems spanning an
enormous range of compact object mass: from x-ray binaries (XRBs) and
gamma-ray bursts (GRBs) with neutron stars and stellar-mass black
holes (BHs) of mass $M\sim1.4{-}20M_\odot$, to active galactic nuclei
(AGN) with BHs of mass $M\sim10^6{-}10^{10}M_\odot$.  The physics of jet production
appears to be robust and insensitive to the
details of the central object.

Two characteristic features of relativistic jets are the following:
(i) Jets are accelerated efficiently and achieve large Lorentz factors
$\gamma$, ranging to above several hundred in the case of GRBs.  (ii) Jets are
strongly confined, with opening angles $\Theta_j\lesssim0.1$~rad.
It is generally assumed that these two properties are
related, though the exact relation is not well understood.

Long-duration GRBs (which we hereafter refer to simply as GRBs) are
particularly interesting in this regard since we have measurements of both
$\gamma$ and $\Theta_j$
\citep{pir05}.
Many GRBs have $\gamma\gtrsim400$ \citep{lithwick_lower_limits_2001},
with jet opening angles broadly distributed
around a typical value $\Theta_j\sim0.05$~rad
\citep{zeh_long_grb_angles_2006}.  These estimates imply that
$\gamma\Theta_j\sim10{-}30$.

Jet acceleration requires an energy source.  In the absence of
magnetic fields this source may be thermal energy injected into the jet by
annihilating neutrinos from an accretion disk
\citep{kohri_neutrino_dominated_2005,chen_neutrino_cooled_2007,kawanaka_neutrino_cooled_2007,zb09}.
Such thermally-driven jets accelerate because of
expansion of the relativistically hot gas
\citep[e.g.,][]{zhang_relativistic_jets_2003,zwh04,mor06,
  wang_relativistic_2007}.
However, self-consistent simulations including neutrino physics
have not yet succeeded in launching relativistic jets
(\citealt{nagataki_neutrino_jets_2007,takiwaki_special_2007,nagataki_grb_sn09}; however,
see \citealt{hkt10}).
The inclusion of magnetic fields
enables the extraction of rotational energy from the central compact
object and this naturally produces relativistic jets
(\citealt{bz77,mck04,mck06jf,km07,tch08}, hereafter \citetalias{tch08}).
In such models, 
acceleration is caused by the expansion of magnetic fields which have been
twisted and amplified by the
rotation of the central compact object
(\citealt{begelman_asymptotic_1994,kom09}, hereafter
\citetalias{kom09}; \citealt{tch09}, hereafter \citetalias{tch09}).
We concentrate on this magnetic acceleration mechanism in the
present paper.

\citetalias{kom09} argued, based on numerical simulations, that relativistic magnetohydrodynamic (MHD)
jets confined by an external medium should have
$\gamma\Theta_j\lesssim1$.  Their result
appears to be in conflict with the GRB data summarized above.
On the
other hand, a completely unconfined split-monopole flow achieves
larger opening angles and gives $\gamma\Theta_j>10$ \citepalias{tch09}.
How do these results relate to GRBs?
According to the collapsar model \citep{woosley_gamma_ray_bursts_1993,mac99},
GRB jets are produced by an
accreting stellar-mass black hole at the center of a collapsing massive
star.  As the jet propagates out, it is collimated by the pressure of
the shocked stellar envelope.  However, once the jet emerges from the star it
is effectively in vacuum, since the pressure of the external
interstellar medium or
stellar wind is far lower than the internal pressure in the jet.
Thus a GRB jet
corresponds to a hybrid scenario which involves initial confinement followed
by free propagation.  What are the asymptotic properties of such a jet
far from the star?  In particular, does $\gamma\Theta_j$ have a value
of order unity, as in confined models \citepalias{kom09},
or is it comparable to the values seen in unconfined split-monopole
models \citepalias{tch09}?

We have carried out high-resolution
simulations that model the behavior of a relativistic
magnetized jet emerging from stellar confinement.
Our ideal MHD jets are confined out to a
certain distance by a rigid wall of the appropriate shape to mimic the
confining effect of the shocked stellar envelope (see
\S\ref{sec_nummethod}).  We then allow the jets to be effectively
unconfined beyond this distance.  The simulations are meant to represent
quasi-steady GRB jets, i.e., we imagine that an initial (possibly non-relativistic) precursor
jet \citep[e.g.,][]{burrows_snjets_2007} has already cleared the way
for an ultrarelativistic magnetized jet which later powers the GRB.
It is the latter jet that our simulations are meant to represent.
The numerical setup is discussed in
\S\ref{sec_nummethod} and the results are described in
\S\ref{sec_results}.  We conclude in \S\ref{sec_discussion} with a
discussion of the results in the context of GRBs.  We work
with Heaviside-Lorentzian units, neglect gravity, and set the speed of light
and the compact object radius to unity: $c=r_0=1$. For a
maximally-spinning BH with a characteristic mass $M=3M_\odot$ \citep{mac99}, our unit
of length, in physical units, is
$GM/c^2 \approx 4.4\times 10^5$~cm.

%%%%%%%%%%%%%%%%%%%%%%%%%%%%%%%%%%%%
%%
%% FIGURE 1: 2D JET
\label{sec_setup}
\begin{figure*}[t!]
  \begin{center}
      \includegraphics[width=\figwidthfactor\textwidth]{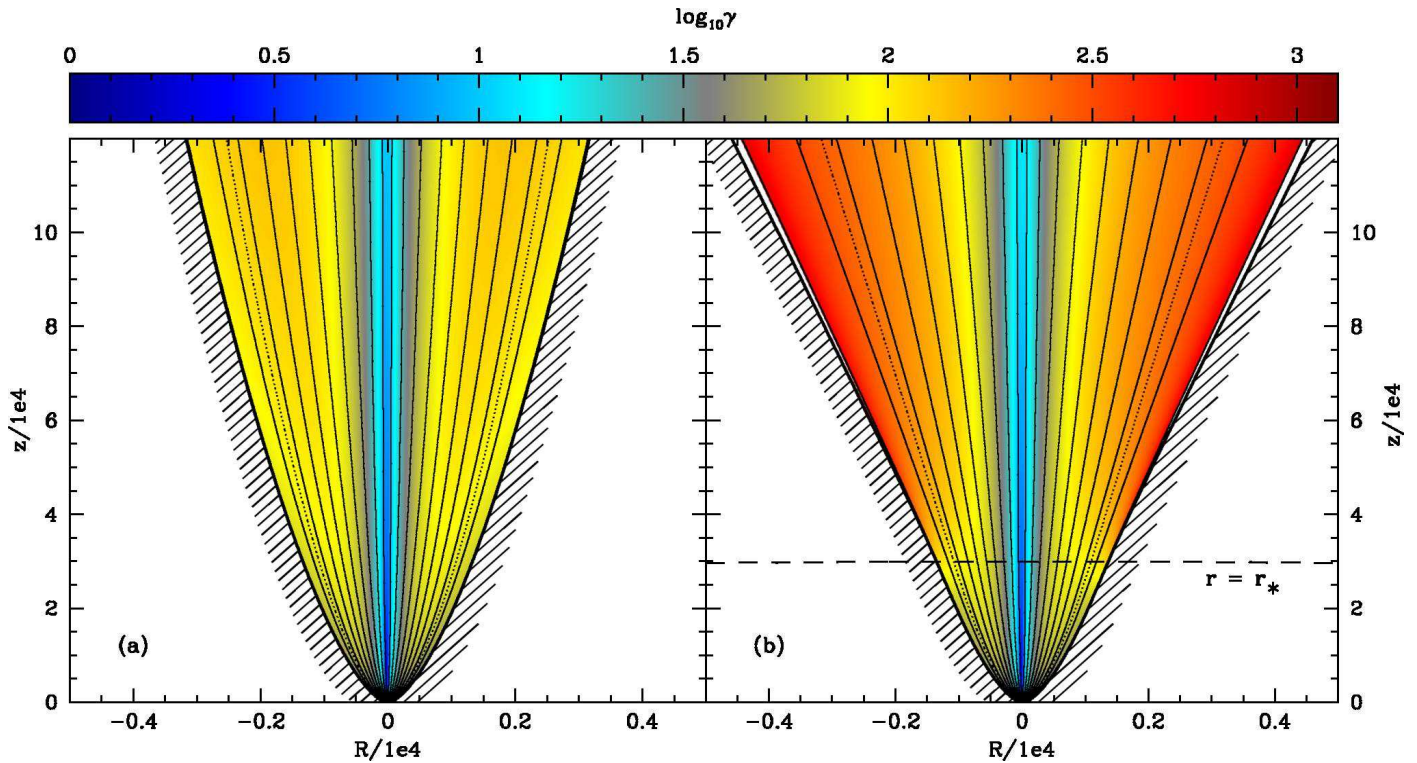}%
  \end{center}
  \caption{Meridional jet cross-section, showing color-coded logarithm
    of the Lorentz factor $\gamma$ overlaid with poloidal field lines
    (thin solid
    lines corresponding to $\Psi^{1/2}=0.1,0.2,\dots,1$). Thick solid
    lines show the position of the wall and dotted lines show the
    half-energy field line.   [Panel a]: Jet
    confined continuously by a wall out to a large distance (model M$\infty$).
    [Panel b]: Jet
    confined by a collimating wall until $r=r_*=3\times10^4$ (model M4, where
    $r_*$ is shown by the horizontal dashed line),
    after which the wall opens up and the jet becomes deconfined.  At this point, the jet
    separates from the  wall and is surrounded effectively by vacuum.
    Once the jet becomes deconfined, $\gamma$
    increases abruptly and the jet opening angle increases slightly.}
  \label{fig_1}
\end{figure*}
%% END FIGURE 1
%%
%%%%%%%%%%%%%%%%%%%%%%%%%%%%%%%%%%%%

\section{Numerical Method and Problem Setup}
\label{sec_nummethod}

We have performed our simulations using the relativistic MHD code HARM
\citep{gam03,mck04,mck06jf}, including recent improvements
(\citealt{mm07,tch07}; \citetalias{tch08}; \citetalias{tch09}).
We initialize the simulation with a purely
poloidal field configuration in which the shape of each field line is
given in polar coordinates
by\begin{equation}\label{eq:thetaff}\theta(r)\propto{}r^{-\nu/2},\quad0.5\lesssim\nu\lesssim1.\end{equation}
\citetalias{tch08} showed that the above power-law shape
corresponds to a power-law profile for the
confining pressure:
\begin{equation}
  \label{eq:alpha}
  p\propto r^{-\alpha}, \quad \alpha = 2(2-\nu) .
\end{equation}
We discuss our choice of $\nu$ in  \S\ref{sec_results}.
In the simulation, the jet shape~\eqref{eq:thetaff}
is maintained until a distance $r=r_*$, where $r_*$ represents
the radius of the confining stellar envelope.  Beyond $r_*$, we allow
the field to decollimate.

In order to arrange for a smooth transition
at $r_*$, we suitably choose the flux function $\Psi(r, \theta)$
of the initial magnetic field,
\begin{equation}\label{eq:psi1}\Psi(r,\theta)=(1-\cos\theta)/f^2(r),\end{equation}where
we note that, by definition, $\Psi$ remains constant on field lines, and the total
poloidal magnetic flux  enclosed inside a polar angle
$\theta$ at radius $r$ equals $2\pi\Psi(r,\theta)$.  The function $f(r)$
essentially sets the opening angle of the wall and we adjust this
function so as to
match our requirements.  Inside the star, where the
stellar envelope actively collimates the jet, we choose $f$ such that it
decreases with distance as $f=(r/r_{\rm
  in})^{-\nu/2}$, giving the collimating shape~\eqref{eq:thetaff}.
Just outside the star, where pressure confinement ceases, we allow the wall
to start decollimating.  To model the deconfinement
we choose $f(r)$ such that it reaches a minimum at $r \simeq r_*$ and
then, within a radial distance $\Delta{}r=0.5r_*$, smoothly
transitions to a logarithmic increase with increasing $r$. The exact
manner in which the wall opens up is unimportant as we discuss in
\S\ref{sec_results}.

The code uses internal coordinates $(x_1,\medspace{}x_2$), which are
uniformly sampled with $1536\times256$ grid cells.  These are mapped to
the physical coordinates ($r,\medspace{}\theta$) via
$r=2.1+\exp(x_1)$, and $x_2={\rm{}sign}(\Psi)\abs{\Psi}^{1/2}$.  In
the lateral $\theta$ direction, the computational domain extends from
the polar axis $x_2=0$, where we apply antisymmetric boundary
conditions, to $x_2=1$, where we place an impenetrable wall whose
shape is determined by equation \eqref{eq:psi1} with $\Psi=1$.
In the radial direction, the computational domain extends from
$r_{\rm{}in}=3$ (three radii of the compact star\footnote{It is a
  matter of convenience where we locate the inner surface at which we
  inject the jet.  We have verified that the solution is insensitive
  to the choice if the flow is sub-Alfv\'enic at this
  surface.}) to $r_{\rm{}out}=3\times10^{11}$.  At $r=r_{\rm{}out}$ we
apply standard outflow boundary conditions, while at $r=r_{\rm{}in}$ we
apply jet injection boundary conditions \citepalias{tch09}.
In detail, at $r_{\rm in}$ we set the
jet magnetic flux  by specifying the radial magnetic field
component according to~\eqref{eq:psi1}, $B_{r,0}=r_{\rm{}in}^{-2}$,
and set the mass flux $j_{\rm{}MA}$ into the jet by specifying the plasma
density $\rho_0=0.00132 B_{r,0}^2$ ($\approx1.2\times10^5$~g~cm$^{-3}$
in cgs units for $B_{r,0}=10^{15}$~G) and poloidal $3$-velocity $v_0=0.5$, both of
which are independent of $\theta$.

We choose the rotational frequency of the central compact star to be
$\Omega=0.25$, which corresponds to a maximally spinning BH,
although
we expect similar results for other values of $\Omega$
\citep{gammie_bh_spin_evolution_2004,mck05}.
We start the
simulation with a non-rotating star and smoothly turn on the rotation.
This generates a set of outgoing waves as the initial purely poloidal
magnetic field develops a helical structure in response to the
rotation.
In the steady-state solution, which quickly establishes behind these waves and
which we study below, energy flowing out of the
compact star is the sum of (a) Poynting flux $j_{\rm{}EM}$,
set by the rotation rate $\Omega$ and the field strength
$B_{r,0}$, and (b) kinetic energy flux $j_{\rm{}KE}$,
set by $\rho_0$ and $v_0$. Poynting flux is converted to kinetic
energy as plasma accelerates along a field line and their ratio
defines the local magnetization of the field
line, $\sigma={j_{\rm{}EM}}/{j_{\rm{}KE}}$.
As
the field line accelerates, magnetization $\sigma$ decreases and
Lorentz factor $\gamma$ increases in such a way that $\mu$, the ratio of the
total energy flux $j_{\rm{}EM}+j_{\rm{}KE}$ to the mass flux $j_{\rm{}MA}$, is conserved,
i.e.,\begin{equation}\label{eq:mu}\mu=\frac{j_{\rm{}EM}+j_{\rm{}KE}}{j_{\rm{}MA}}=\gamma(\sigma+1)={\rm{}constant\thickspace{}along\thickspace{}each\thickspace{}field\thickspace{}line},\end{equation}where $j_{\rm{}KE}=\gamma{}j_{\rm{}MA}$.  Since $\sigma\ge0$, the
value of $\mu$ determines the maximum possible Lorentz factor of the
field line.  The Poynting flux vanishes at the jet axis, so
$\mu(\theta=0)=\gamma_0=(1-v_0^2)^{-1/2}$, and $\mu$ increases away from
the axis roughly proportional to $\Psi$ \citepalias{tch09}, reaching
its maximum value, $\mu_0\approx1150$ (for our choice of boundary conditions),
at the jet boundary.

In the context of GRBs, we are interested in a jet which
asymptotically has a Lorentz factor of order a few hundred.  Also,
we assume that the jet energy
goes into accelerated particles, and thereby to prompt GRB
radiation, via an MHD shock.
In order to have efficient conversion of jet energy into thermal
energy at the shock, it is known that
the jet fluid must achieve $\sigma \lesssim1$ just prior
to the shock \citep{kennel_confinement_crab_1984}.  These
conditions require $\mu\sim10^3$ and motivate our choice of
$\rho_0\sim10^{-3}$.  By choosing this value of $\mu$, we
\emph{design} our simulation such that, if the jet achieves
a Lorentz factor of a few hundred, it will have $\sigma \lesssim 1$.

\begin{figure}[t!]
\begin{center}
\includegraphics[width=\columnwidth]{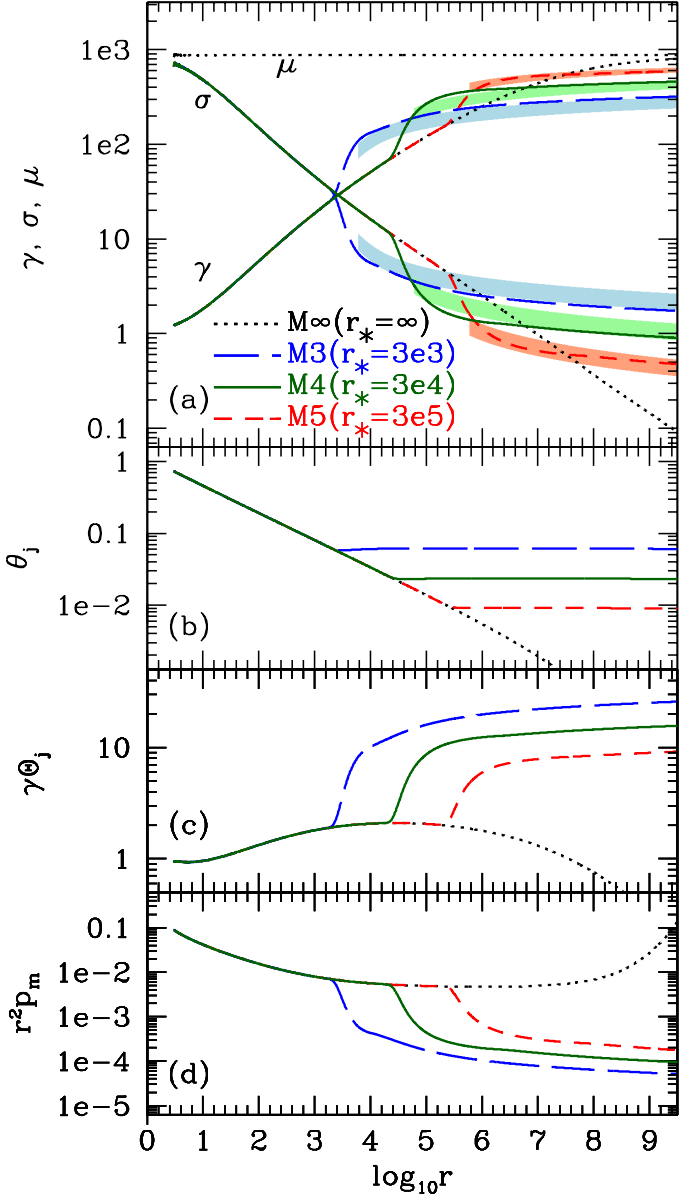}
\end{center}
\caption{Dependence
    of various quantities along the half-energy field line for models
    M$\infty$, M$3$, M$4$, and M$5$.  These models differ only in the
    jet confinement radius, i.e., the radius of the star $r_*$.
    [Panel a]: Lorentz factor $\gamma$, magnetization $\sigma$, and
    the total energy flux $\mu$. Model M$\infty$ shows continuous
    smooth acceleration, whereas the deconfined models M$3$--M$5$ show
    a ``burst'' of acceleration soon after the jet loses
    confinement. Shown with color stripes is the analytic solution for
    a fully unconfined jet (eq.~\ref{eq:gamma_anal}).  It reproduces our numerical solutions to
    within $20$\% (the width of the stripes).  [Panel b]: Field line
    opening angle $\theta_j$.  This decreases continuously while the jet is
    confined by the wall, but it hardly changes outside the star.
    [Panel c]: The product of the field line $\gamma$ and the \emph{full}
    jet opening angle $\Theta_j$.  In model M$\infty$, we have
    $\gamma\Theta_j\lesssim1$ at all $r$.  However, in models
    M$3$--M$5$, $\gamma\Theta_j$ increases abruptly soon after the jet
    becomes unconfined and reaches values $\sim10{-}30$, as
    appropriate for GRB jets. [Panel d]: Magnetic pressure
    $p_{\rm{}m}$.  This decreases smoothly with increasing $r$
    so long as the jet is confined ($r\lesssim{}r_*$).
    However, once the wall opens up, $p_{\rm{}m}$ drops abruptly.}
   \label{fig_2}
\end{figure}

\section{Numerical Results}
\label{sec_results}

We have run several simulations with the parameter $\nu$ set to
$3/4$, and with different values of the confinement radius $r_*$
covering the likely range of progenitor star radii in GRBs.
\begin{figure}[t!]
  \begin{center}
    \includegraphics[width=0.85\columnwidth]{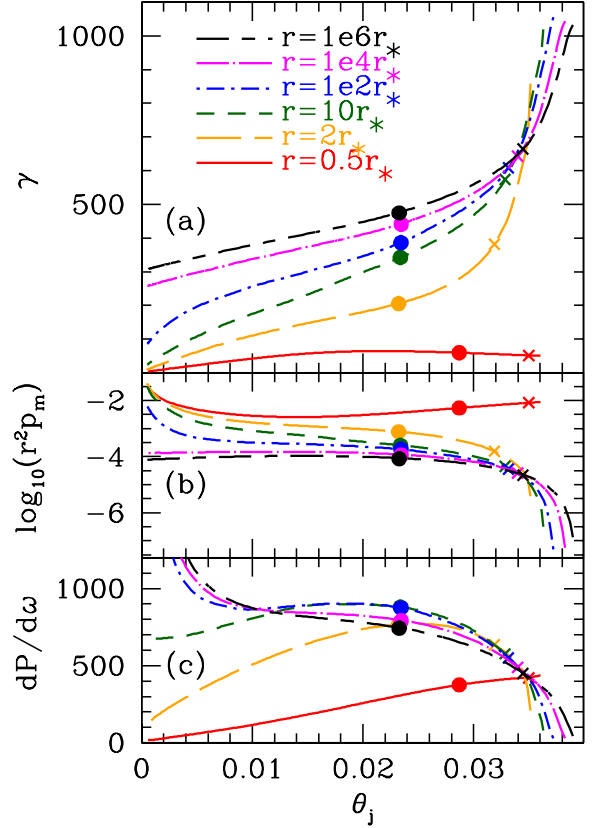}
  \end{center}
  \caption{Angular distribution of $\gamma$ (panel a), magnetic
    pressure (panel b), and power output per unit solid angle (panel
    c) in our fiducial model M$4$ as a function of $\theta_j$, shown
    at various distances (see legend). Thick dots indicate the
    position of the half-energy field line.  The collimation angle of
    this field line is nearly constant, $\theta_j\approx0.024$, once the jet becomes
    deconfined.  However, its Lorentz factor
    increases dramatically.  Crosses indicate the full jet opening
    angle $\Theta_j$ (\S\ref{sec_results}).
    Once the jet leaves the star,
    magnetic pressure and power output along the outermost field lines
    drop essentially to zero.  Thus, outside the star, the jet is
    surrounded by a vacuum.}
  \label{fig_3}
\end{figure}The
choice $\nu=3/4$ is motivated by the work of \citet{mck07a,mck07b}
and especially \citetalias{tch08}, who showed that this value
corresponds to a confining pressure profile, $p\propto{}r^{-5/2}$
(eq. \ref{eq:alpha}), which is
reasonable for a jet-shocked stellar envelope, e.g., see
hydrodynamic simulations of
relativistic jets injected into GRB progenitor stars by
\citet{zhang_relativistic_jets_2003}.
We have also run models with $\nu=2/3$ (i.e., $p\propto{}r^{-8/3}$),
and the results are not
qualitatively different.

As a baseline model, we first consider the case $r_*\to\infty$, i.e.,
a model in which the jet is continuously collimated by a wall out to
an arbitrarily large distance. We refer to this as model M$\infty$.
Figure~\ref{fig_1}a shows a meridional cut through the steady-state
solution we obtain at the end of the simulation.  In this model, the jet
$\gamma$ increases steadily and continuously and becomes
of order a few hundred at large distance, while the jet opening angle
decreases continuously with increasing distance.

Consider next a model with $r_*=3\times10^4$ ($=1.3\times10^{10}$~cm
for a $3M_\odot$ black hole).  In this model, which we refer to as model M4,
the wall collimates exactly as in model M$\infty$ until $r\sim{}r_*$,
but the wall then smoothly opens up (as described by
eq.~\ref{eq:psi1}), allowing the jet to decollimate.
The steady state solution is shown in
Figure~\ref{fig_1}b.  Just beyond $r=r_*$,
soon after the jet loses confinement, we see that (a) the Lorentz factor $\gamma$
increases abruptly by a factor $\sim10$ and (b) the opening angle of
the jet $\Theta_j$ shows a slight increase and then
``freezes out,'' hardly changing thereafter.

Figure~\ref{fig_2} shows the radial dependence of various quantities.
In addition to the models M$\infty$ and M4 already discussed, we
consider two additional models: model M3 with $r_*=3\times10^3$
($1.3\times10^9$~cm) and
model M5 with $r_*=3\times10^5$ ($1.3\times10^{11}$~cm).
For each model, we show results
corresponding to the ``half-power'' field line, i.e., the field line
for which half the jet power is carried by field lines inside of this
line and half by field lines outside.

Figure~\ref{fig_2}a shows as a function of $r$ the total energy flux
$\mu$ (which is constant and the same for all models), the Lorentz
factor $\gamma$, and the magnetization parameter $\sigma$.  While
model M$\infty$ shows continuous smooth acceleration out to large
radius, each of the three deconfined models, M3, M4 and M5, exhibits an abrupt
increase in $\gamma$ soon after the jet loses confinement, with most
of the acceleration occurring between about $r_*$ and $2r_*$.

Figure~\ref{fig_2}b shows the dependence of the angle $\theta_j$
between the poloidal component of the half-power field line and the
jet axis. We use $\theta_j$ for the opening angle of
a given field line and $\Theta_j$ for the opening angle of the
entire jet; for the latter we use the field line that
contains $90$\% of the jet's power output.  Figure~\ref{fig_2}b shows that,
so long as the jet is confined
($r\lesssim{}r_*$),
$\theta_j$ decreases with distance as $r^{-3/8}$
(eq. \ref{eq:thetaff} with $\nu=3/4$).  However, once the jet crosses $r_*$, the
opening angle of the field line freezes and remains nearly constant
out to arbitrarily large distance.
Therefore, it is the opening angle of the wall
(eq.~\ref{eq:psi1}) as evaluated at $r=r_*$ that determines
the asymptotic value of the jet opening angle.  We note that the opening angles
of our jets are similar to those found in hydrodynamic
simulations of GRB jets \citep[e.g.,][]{zhang_relativistic_jets_2003}.
This is perhaps not surprising since we calibrate the shape
of the confining wall in our models by the shape of the jet boundary in
the hydrodynamic simulations.

After the burst of acceleration between $r_*$ and $2r_*$, our deconfined jets
subsequently accelerate
very weakly. This is similar to unconfined
magnetically-accelerated outflows which are known to
accelerate only logarithmically at large distance (\citealt{bes98}, \citetalias{tch09}).
This is in contrast to thermally-accelerated outflows which continue to accelerate efficiently
until they reach their maximum Lorentz factor
\citep[e.g.,][]{zhang_relativistic_jets_2003,mor06}.

Once an initially confined jet is allowed to become
deconfined, its structure rearranges and begins to resemble
a fully unconfined jet.
Figure~\ref{fig_2}a shows
as colored stripes the approximate analytic solution for
unconfined outflows obtained by \citetalias{tch09}:
\begin{equation}
  \label{eq:gamma_anal}
  \gamma=2\left(\frac{\mu-\gamma}{\theta_j^2}\log{}\frac{\Omega{}R}{\gamma_c}\right)^{1/3},
  \quad{}\gamma_c=\left(\frac{\mu}{\sin^2\theta_j}\right)^{1/3}.
\end{equation}
Here
$R=r\sin\theta_j$ is the cylindrical radius, $\log$ is the natural logarithm, and
$\theta_j=(1-\nu/2)(2\Psi)^{1/2}(r_*/r_{\rm{}in})^{-\nu/2}$ is the angle of the
half-power field line, evaluated at the stellar surface
(equation~\eqref{eq:psi1} with
$\Psi\approx0.7\thickspace{\rm{}and}\thickspace{}f=(r_*/r_{\rm{}in})^{-\nu/2}$). The
agreement of the analytical solution with our numerical results is better than $20$\%.
This agreement suggests that, once
deconfined, our jets rapidly relax to the equivalent
unconfined monopole solution for the same angle $\theta_j$ by converting
their excess magnetic energy into kinetic energy.
There is, however, an important difference.  In a pure
monopole solution, most of the jet energy flows out in the equatorial region,
whereas in our jet models, because of the prior collimation,
the energy flow is confined to within
$0 \leq\theta_j\leq\Theta_j$.

Figure~\ref{fig_2}c shows the radial variation of the product
$\gamma\Theta_j$, where $\Theta_j$ is the angle of the jet boundary
(the field line with $\Psi=0.95$, which encloses $90$\% of the jet's
total power).
For the
continuously collimated jet in model M$\infty$ (dotted line),
$\gamma\Theta_j$ is of order unity throughout the acceleration zone,
as noted by \citetalias{kom09}, decreasing to a much smaller value at
large distance.  The unconfined jets in M3, M4 and M5, however, behave
very differently.  Once these jets become unconfined, $\gamma\Theta_j$
increases rapidly from $\simeq1$ to $\sim10{-}30$.  An angular
power-density--weighted average of $\gamma\theta_j$ gives values
within $30$\% of this estimate, which shows that the details of how we
select $\gamma$ and $\Theta_j$ are unimportant. (A jet break is
expected only if a large fraction of the power has
$\gamma\theta\gg 1$; the power-density weighting accounts for this.)

What determines the value of $\gamma\Theta_j$ for a jet?  To find this
out, let us rewrite equation \eqref{eq:gamma_anal} in terms of
magnetization, using equation \eqref{eq:mu}:
\begin{equation}
  \label{eq:gammathetaanal}
  \gamma\Theta_j = 2^{3/2} (\Theta_j/\theta_j)\sigma^{1/2} \log^{1/2}\frac{\Omega R}{\gamma_c}
  \simeq 15 \sigma^{1/2},
\end{equation}
where in the last equality we assumed characteristic values for jet
parameters in model M4 ($r=10^9$, $\theta_j=0.02$, $\Omega=0.25$, $\mu
= 10^3$, $\Theta_j/\theta_j=1.5$).  From this equation we clearly see that jets with
subdominant magnetic fields at large distance, $\sigma\lesssim1$, are limited to
$\gamma\Theta_j\lesssim 15$.

We note that higher values of $\gamma\Theta_j$ correspond to
deconfined models with smaller values of
$r_*$.  From the simulation results shown in Figure~\ref{fig_2}c,
we obtain a scaling
\begin{equation}\label{eq:gthscalingnum}\gamma\Theta_j\propto{}r_*^{-0.22}.\end{equation}
We can use the analytic solution of
fully unconfined jets to understand this trend.
As we show in the Appendix, for models M3--M5, which are asymptotically
roughly in equipartition ($\gamma\approx\mu/2$ at large distance), the
analytic solution \eqref{eq:gamma_anal} gives:
\begin{equation}
  \label{eq:analytic_scaling2}
  \gamma\simeq\mu/2:
  \quad\Theta_j\propto{}r_*^{-\nu/2},
  \quad\gamma\propto{}\mu^{1/2}r_*^{\nu/4},
  \quad\gamma\Theta_j\propto{}\mu^{1/2}r_*^{-\nu/4}.
\end{equation}
This
implies $\gamma\Theta_j\propto{}r_*^{-0.19}$ for
$\nu=3/4$, in good agreement with the numerical
scaling~\eqref{eq:gthscalingnum}.  In the limit of high
magnetization, i.e., $\sigma\gg1$ and $\gamma\ll\mu$,  we obtain
instead (see the Appendix)
\begin{equation}
  \label{eq:analytic_scaling}
  \gamma\ll\mu:
  \quad\Theta_j\propto{}r_*^{-\nu/2},
  \quad\gamma\propto\mu^{1/3}r_*^{\nu/3},
  \quad\gamma\Theta_j\propto\mu^{1/3}r_*^{-\nu/6}.
\end{equation}
Therefore,
smaller values of stellar radius $r_*$ and larger values of $\mu$ give
larger values of $\gamma\Theta_j$.  However, in this limit, the
asymptotic value of $\sigma$ will be $\gg1$.  Such jets cannot
efficiently convert their Poynting energy flux into particle thermal energy
via an MHD shock and therefore it is not clear if they will have sufficient
energy in accelerated electrons to produce the prompt $\gamma$-ray emission
observed in GRBs
(see end of \S\ref{sec_nummethod}).  Other particle acceleration
mechanisms, 
e.g., magnetic reconnection or plasma instabilities
\citep[e.g.,][]{beloborodov_collisional_grb_10}, or different mechanisms,
e.g., inverse Compton scattering of thermal photons \citep[e.g.,][]{bro05},
may be able to circumvent this limit.

Finally, Figure~\ref{fig_2}d shows the profile of magnetic pressure
along the half-power field line.  In the confined region of the flow,
the pressure varies roughly as $r^{-5/2}$, leveling off to
$\sim{}r^{-2}$ at larger distance.\footnote{The pressure profile in
  the slower magnetized sheath that surrounds the jet in a real system
  continues to follow  the
  $p\propto r^{-5/2}$ dependence~\citepalias{tch08}.}  However, as
the jet emerges from confinement, poloidal field lines
rarefy laterally and the pressure decreases by over a factor of $10$
(see models M3--M5).  The drop in pressure creates a strong
longitudinal pressure gradient which is the reason for the abrupt
acceleration of the jet.  This is an extreme example of the ``magnetic nozzle''
effect  (\citealt{begelman_asymptotic_1994}; \citetalias{tch09}) in
action.

Figures \ref{fig_3}a,b,c show for model M4 the transverse profiles of
$\gamma$, magnetic pressure $p_{\rm{}m}$, and power output per unit
solid angle $dP/d\omega$ at various distances.  We see that the abrupt
acceleration and pressure drop experienced by the half-power field
line (indicated by thick dots), described earlier in
Figures~\ref{fig_2}a,d, is representative of most field lines.  The
outermost field lines near the edge of the jet ($\theta_j>\Theta_j$)
accelerate enormously during deconfinement but they contain little
power ($\lesssim10$\%) (\citealt{lyub09}).  Along these field lines
the pressure drops essentially to zero once the jet emerges from the
star, so these outer field lines are surrounded by vacuum.

We have included a fairly gentle
opening up of the wall in the simulated models (e.g., see the shape of the
wall in Fig.~\ref{fig_1}b).  However, because the jet moves
ultrarelativistically and all signals are strongly beamed forward,
even small changes in geometry are equivalent to nearly
complete deconfinement, e.g., the jet surface
in Fig.~\ref{fig_1}b clearly separates from the wall and the pressure
at the jet surface (Fig.~\ref{fig_3}c) goes to zero.
We have confirmed that our results are
insensitive to the precise shape of the opening-up of the
wall.
Since the wall is causally disconnected from the flow, a more strongly
diverging wall shape would not affect the results.

It should be noted that the numerical simulations described in this paper are extremely
challenging since at high magnetization
the equations of motion are very stiff and hard to
solve. Additionally, it is a challenge to obtain accurate
solutions that extend over $10$
orders of magnitude in distance and in which ideal MHD conserved quantities
are preserved 
along field lines to better than $15$\% \citepalias{kom09,tch09}; our
numerical  models achieve this accuracy and are
well-converged.  To ensure that we have
sufficient angular resolution near the pole
and the wall, and sufficient radial resolution to resolve abrupt radial changes of quantities where
the wall opens up, we have rerun model M4 with a locally $5$
times higher angular resolution near the pole and near the wall, and also
with a $5$ times higher radial resolution between $0.5r_*$ and $4r_*$.
We obtained less than $2$\% difference in asymptotic values of
$\gamma$ and $\theta_j$ for most of the jet cross-section
($0.2<\theta_j/\Theta_j<1$).  After this work was posted on the archives,
our results were confirmed by \citet{kvk09} using a completely independent code.

\section{Discussion}
\label{sec_discussion}

GRB afterglow radiation is produced when a relativistic jet ejected
from a collapsar interacts with an external medium and decelerates.
The afterglow phase typically starts at the end of the prompt
gamma-ray emission, say at a time $t_0\sim30\medspace$s in the observer
frame.  The Lorentz factor $\gamma$ and jet opening angle
$\Theta_j$ discussed in the previous sections refer to the jet properties
at this time.  During the subsequent afterglow phase,
the Lorentz factor $\gamma$ of the jet
decreases with time.  In the case of a uniform external medium
and adiabatic evolution, $\gamma$ varies with expansion distance
as $r^{-3/2}$ \citep{bm76}, which corresponds
to a variation with observed time as $t^{-3/8}$
(e.g., \citealt{spn98,pir05}).  The opening angle of the jet
$\Theta_j$, however, does not change.  Therefore, during the afterglow
phase of a GRB, the quantity $\gamma\Theta_j$ varies with observer time as
\begin{equation}
  \gamma\Theta_j\approx(t/t_0)^{-3/8}(\gamma\Theta_j)_0,
  \label{eq:gammatheta}
\end{equation}
where the subscript $0$ indicates values measured at the end of the
prompt emission stage.
When $\gamma\Theta_j$ falls below
unity, there is an achromatic break in the afterglow light-curve and
the observed flux falls more steeply with time
\citep{rhoads_uniform_jet_1999,sph99}.  This ``jet break'' has been
seen in a number of GRBs, typically about one to ten days after the initial
prompt emission \citep{frail01,cenko_grb_breaks_2009}.
For these GRBs, according to
equation (\ref{eq:gammatheta}), we require an initial
$(\gamma\Theta_j)_0\sim(1\thickspace{\rm{}day}/30\thickspace{\rm{}s})^{3/8}\sim20$
to explain the observations.  This constraint has to be satisfied by
any model of a GRB jet.

We find that both confinement by the collapsar's stellar envelope
and free propagation outside the star are needed for an MHD
jet to have (i) the required value of $(\gamma\Theta_j)_0$, (ii) the
large power inferred in long-duration GRBs, and (iii) efficient
conversion of electromagnetic to kinetic energy ($\sigma\lesssim1$ at
large distance).  Purely unconfined jets do give
$(\gamma\Theta_j)_0\gg1$ \citepalias{tch09}.  However, they have $\sigma\gg1$ over
most of their volume, which is undesirable, and they have too
little power within the relevant opening angle where $\sigma\lesssim1$.  Purely confined jets are much
better at focusing the power of the central engine into a collimated
beam with $\sigma\lesssim1$, but they give $(\gamma\Theta_j)_0\lesssim1$.  Only jets that
are first confined and then deconfined can explain the most energetic
long-duration GRBs with achromatic jet breaks.  Such jets
abruptly accelerate as soon as they emerge from the stellar
envelope and develop a typical value of $(\gamma\Theta_j)_0\sim20$,
while at the same time achieving $\sigma\lesssim1$ (Fig.~\ref{fig_2}).
Remarkably,
the stellar confinement radius we need to produce the observed value of
$(\gamma\Theta_j)_0$ is $r_*\sim 10^{4-5}$ or a physical
radius $\sim{\rm few}\times10^{10}$\ cm,
which agrees well with the estimated radii of GRB
progenitor stars \citep[e.g.,][]{hl00}.

Achromatic jet breaks have been observed in
$20$\% of the \emph{Swift} GRBs, but the actual number is much higher
($\sim50$\%) since  observations of
most GRBs ceased before the expected jet break time \citep{kb08}.
As we have shown, current observations are putting stringent constraints on
GRB models.
If future observations by the Fermi Observatory routinely find $\gamma_0>10^3$ \citep{kd09} and $(\Theta_j)_0 > 0.1$
\citep{cenko_grb_breaks_2009} for many GRBs, we
would need to explain why these jets have $(\gamma\Theta_j)_0>100$.
The numerical factor in equation \eqref{eq:gammathetaanal} is a
logarithmic factor and is unlikely to be larger than $10{-}20$. Therefore,
either we must accept that $\sigma\gg1$, i.e., GRB jets are
Poynting-dominated and somehow manage to convert a large fraction of
their energy to prompt gamma-rays, or that MHD is not the appropriate
framework for understanding GRB jets.

While we have focused primarily on GRB jets, our results also have
some bearing on relativistic jets in XRBs and AGN.  In order to
produce collimated and efficient ($\sigma\lesssim1$) jets,
the accretion disks in these systems must collimate the
outflowing Poynting flux from the central BH. Strong winds are
naturally present in advection-dominated accretion flows (ADAFs,
\citealt{nar94,nar95a,nm08}), so one expects to see well-developed jets
whenever XRBs and AGN are in the ADAF state.  The exact properties of the
emerging jet will presumably depend on the distance to which efficient
wind-collimation operates.

We note that our simulations assume a fixed shape for the
confining wall.  However, as the
GRB progresses and the jet disrupts the star, the opening
angle of the jet is likely to increase.  Our model does not take this
effect into account.  Also, our assumption of a rigid wall as the jet
confining agent eliminates instabilities and mixing at the jet
interface. These effects should be taken into account in future work,
although the problem of a magnetized ultra-relativistic jet interacting
with a surrounding progenitor star envelope is numerically
extremely challenging and remains unsolved.
Future
work should also study jet stability to
non-axisymmetric modes \citep{nlt09}, 
general relativistic effects of the central spinning BH \citep{tch10},
how GRB engines
become threaded by the ordered magnetic fields needed for MHD jets
\citep{mck06jf,mb09}, and should model radiation processes
in order to make detailed comparisons with observed afterglow
light curves and jet breaks.

\acknowledgements
We thank the anonymous referee for valuable comments on the
manuscript.
This work was supported in part by NASA grant
NNX08AH32G (AT \& RN), NSF grant AST-0805832 (AT \& RN),
NASA Chandra Fellowship PF7-80048 (JCM), and by
NSF through TeraGrid resources~\citep{catlett2007tao}
provided by the Louisiana Optical Network Initiative
(\href{www.loni.org}{www.loni.org}) under grant number TG-AST080026N.

% \bibliography{mybib}
% \ifthenelse{\equal{\useiop}{1}}{
%   % IOP
%   \bibliographystyle{jphysicsB} }

% \ifthenelse{\equal{\useiop}{0}}{
%   % APJ
%   \bibliographystyle{apj}
% }

% \ifthenelse{\equal{\useiop}{-1}}{
%   % APJ
%   \bibliographystyle{apj}
% }

% \ifthenelse{\equal{\useiop}{-2}}{
%   % APJ
%   %\bibliographystyle{elsarticle-harv}
%   \bibliographystyle{apjrepeat}
% }

\appendix
\section{Analytic scaling of $\gamma\Theta_j$ in jets}
In this section, we analytically derive relations
for $\gamma\Theta_j$. For this, we substitute $\Theta_j
\propto r_*^{-\nu/2}$ (see eq.~\ref{eq:thetaff}) into equation
(\ref{eq:gamma_anal}) and neglect the logarithm:
\begin{equation}
  \label{eq:gamma_equipartitionapp}
  \frac{\gamma^3}{\mu-\gamma}\propto r_*^\nu.
\end{equation}
Now, we take a logarithm of both sides of equation
\eqref{eq:gamma_equipartitionapp} and take a differential:
\begin{equation}
\label{eq:gamma_generalapp}
3\frac{d\gamma}{\gamma} - \frac{d\mu-d\gamma}{\mu-\gamma} =
\nu\frac{dr}{r_*}
\end{equation}
Our models M3--M5 are asymptotically roughly in
equipartition: $\gamma\simeq\mu/2$ at large distance.
Therefore, we substitute $\mu-\gamma=\gamma=\mu/2$ in the result,
\begin{equation}
  \label{eq:gamma_eqipartition1app}
4\frac{d\gamma}{\gamma} - 2\frac{d\mu}{\mu}= \nu\frac{dr}{r_*}.
\end{equation}
Integration of this equation gives:
\begin{equation}
  \label{eq:analytic_scaling2app}
  \gamma\simeq\mu/2:
  \quad\Theta_j\propto{}r_*^{-\nu/2},
  \quad\gamma\propto{}\mu^{1/2}r_*^{\nu/4},
  \quad\gamma\Theta_j\propto{}\mu^{1/2}r_*^{-\nu/4}.
\end{equation}
This
implies $\gamma\Theta_j\propto{}r_*^{-0.19}$ for
$\nu=3/4$, in good agreement with the numerical
scaling~\eqref{eq:gthscalingnum}.  In the limit of high
magnetization, i.e., $\sigma\gg1$ and $\gamma\ll\mu$,  we
approximate $\mu-\gamma\approx\mu$  in equation
\eqref{eq:gamma_equipartitionapp} and obtain instead
\begin{equation}
  \label{eq:analytic_scalingapp}
  \gamma\ll\mu:
  \quad\Theta_j\propto{}r_*^{-\nu/2},
  \quad\gamma\propto\mu^{1/3}r_*^{\nu/3},
  \quad\gamma\Theta_j\propto\mu^{1/3}r_*^{-\nu/6}.
\end{equation}

\label{lastpage}
\end{document}